\documentclass[a4paper,12pt]{article}

\usepackage{amsmath,amssymb,mathtools} 
\usepackage{color}
\usepackage{graphicx}
\usepackage{subfigure}
\usepackage{cite}           
\usepackage{hyperref}            
\usepackage{multirow,makecell}   
\usepackage{textcomp}
\usepackage{wasysym}
\usepackage{setspace}
\usepackage{verbatim}
\usepackage{float}
\usepackage{ulem}
\usepackage[utf8]{inputenc}
\graphicspath{{newFig/}}

\usepackage[text={17.2cm,25.2cm},centering]{geometry}

\numberwithin{equation}{section}

\begin{document}
	\title{\textbf{String tension and Polyakov loop in a rotating background}}

	\author{Jun-Xia Chen$^{1}$\footnote{chenjunxia@mails.ccnu.edu.cn },  ~  Sheng Wang$^{1}$\footnote{shengwang@mails.ccnu.edu.cn } ,  ~ Defu Hou$^{1}$\footnote{houdf@mail.ccnu.edu.cn } , ~     Hai-Cang Ren$^{1,2}$\footnote{renhc@mail.ccnu.edu.cn } }
	\date{}
	
	\maketitle
	
	\vspace{-10mm}
	
	\begin{center}
	{\it
		$^{1}$ Institute of Particle Physics and Key Laboratory of Quark and Lepton Physics (MOS),
		Central China Normal University, Wuhan 430079, China\\ \vspace{1mm}
		
		$^{2}$ School of Physical Science and Technology,  ShanghaiTech University, Shanghai 201210, China\\ \vspace{1mm}			
	}
	\vspace{10mm}
\end{center}

\begin{abstract}

We study the influence of a rotation on the string tension and the temperature of the confinement-deconfinement phase transition of gluodynamics by gauge/gravity duality. We explore two distinct approaches, global transformation and local transformation, to introduce rotation and compare the results. 
It is shown that the string tension extracted from the free energy in the presence of a heavy quarkonium decreases with increasing angular velocity,   while the transition temperature determined by the Polyakov loop increases with increasing angular velocity,   which is in line with lattice simulations.

\end{abstract}

\baselineskip 18pt
\thispagestyle{empty}
\newpage

\tableofcontents

\maketitle

\section{Introduction}\label{sec1}
The phase structure of quantum chromodynamics in rotation has been actively investigated in recent years \cite{Chen:2020ath, Zhao:2022uxc, Wang:2024szr, Braga:2022yfe, Chen:2024jet}. The research interests stem from the off-central collision in RHIC where a large angular momentum is stored in the quark-gluon plasma left behind \cite{STAR:2017ckg}, and from rapidly rotating neutron stars. Among theoretical approaches to explore such a strong coupling system are the lattice simulation \cite{Braguta:2021ucr, Braguta:2021jgn, Yang:2023vsw}, holographic QCD \cite{Li:2023mpv, Chen:2022mhf, Zhu:2024dwx} and the mean field theory of NJL model \cite {Jiang:2016wvv, Zhang:2018ome} as well as other field theoretic investigations \cite{ Rybalka:2018uzh, Fujimoto:2021xix, Chen:2020xsr, Chen:2023cjt, Yang:2023zqe, Buzzegoli:2022omv, BitaghsirFadafan:2008adl, Hou:2021own, Chernodub:2020qah, Giantsos:2022qgq}. For a rotating pure gluon system with the linear velocity on the boundary much lower than the speed of light, the lattice simulations reveal that the deconfinement phase transition temperature increases with the angular velocity $\omega$ for $\omega l<<1$, while existing formulations of holographic QCD without introducing additional degrees of freedom on the boundary all point to decreasing deconfinement phase transition temperature with $\omega$ \cite{Chen:2020ath, Zhao:2022uxc, Wang:2024szr, Braga:2022yfe}. As the lattice simulation reflects the first principle of QCD, one needs to explore the alternative formulations of holographic QCD in rotation to resolve the discrepancy. This is the motivation of our present work.

There has not been a consensus approach to introduce a spatial rotation in a holographic framework. Among the formulations explored in literature are the Schwarzschild-$\rm{AdS}^5$ subject to a local boost \cite{BravoGaete:2017dso, Braga:2023qej, Zhang:2023psy, Chen:2022obe}, and Kerr-$\rm{AdS}^5$ \cite{Hawking:1998kw, Arefeva:2020jvo, Golubtsova:2022ldm,  Golubtsova:2021agl, McInnes:2018mwj}. None of them recovers the rotation metric used in the lattice simulation and effective field theories. The opposite dependence of the deconfinement temperature on the angular velocity was extracted from the local boost approach.

In this note, we shall explore analytically the confinement-deconfinement phase transition of gluodynamics in a rotating frame of the boundary coordinates in the Schwarzschild-$\rm{AdS}^5$ spacetime with a dilaton. The string tension \cite{Andreev:2006ct, Andreev:2006eh} in the confinement phase and the Polyakov loop \cite{Andreev:2009zk, Andreev:2006nw, Kaczmarek:2002mc, Kaczmarek:2003dp, Critelli:2016cvq} as an order parameter of the deconfinement phase transition are examined in two different approaches. One is the global rotating frame \cite{Chen:2023yug} with the same boundary metric used in lattice and effective field theories. The other one is the local rotating frame investigated in the literature \cite{Cai:2022nwq, Zhou:2021sdy}. Instead of numerically solving the Einstein equation coupled with matter fields as in \cite{Chen:2020ath, Zhao:2022uxc, Wang:2024szr, Braga:2022yfe, Chen:2024jet}, we transform the simple soft-wall metric employed in \cite{Andreev:2006ct}. Though the deconfinement phase transition in this formulation is characterized by a sharp crossover instead of a discontinuity in the temperature dependence of the Polyakov loop, the pros of analytical tractability outweigh the cons. We found that string tension extracted from the free energy in the presence of a heavy quarkonium is suppressed by the rotation while the deconfinement temperature increases with the angular velocity. The different conclusion from the literature \cite{Zhao:2022uxc, Chen:2020ath} regarding the Polyakov loop may be attributed to a different world-sheet solution for the Polyakov loop found in this work to address the inhomogeneity and anisotropy in a rotating frame \cite{Chen:2023yug}.

This paper is organized as follows. In section \ref{sec2}, the holographic model is reviewed and extended to the global/local rotating background. The string tension and Polyakov loop under rotation are calculated in sections \ref{sec3} and \ref{sec4}. Section \ref{sec5} concludes the paper. The string tension with arbitrary location and orientation, the string tension extracted from internal energy and some formulas underlying the Polyakov loop are deferred to Appendices A, B and C.   

\section{Rotating frame}
\label{sec2}

We will work on the following background metric of Minkowski signature
\begin{equation}
\label{eq201}
  ds^{2}=\frac{R^{2}}{w^{2}}h\left(-fdt^{2}+l^{2}d\phi^{2}+dl^{2}+dz^{2}+\frac{1}{f}dw^{2}\right),  
\end{equation}
with
\begin{equation}
\label{eq202}
   h=e^{\frac{1}{2}cw^{2}},\quad f(w)=1-\frac{w^{4}}{w_{t}^{4}}, 
\end{equation}
where $(t,\phi,l,z,w)$ are coordinates of $\rm{AdS_5}$ of radius $R$ and $w_t$ is the radial position of the horizon. With $c>0$, the metric Eq.(\ref{eq201}) is conjectured to be the gravity dual of gluodynamics at strong coupling on the boundary, $w\rightarrow 0$, and the warp factor $h$ generates confining heavy quark potential and confinement-deconfinement phase transition \cite{Andreev:2006ct, Andreev:2006nw} \footnote{The metric with positive $c$ is expected to be the gravity dual of a pure Yang-Mills theory without dynamic quarks, the same system simulated on the lattice in \cite{Braguta:2021jgn}, where the linear potential between testing quark-antiquark indicates confinement. The metric with negative $c$, on the other hand, serves as a holographic description of the system with dynamic quarks, where the linear potential cannot be sustained because of quark pairs in the Dirac sea and the confinement is reflected in the Regge behavior of the hadron spectrum \cite{Karch:2006pv}.}. The former is characterized by a string tension and the latter is characterized by the nonzero thermal expectation value of a Polyakov loop. We set $c=0.9$ $\rm{GeV^2}$ in this work \cite{Andreev:2006vy}. 

To study the string tension and Polyakov loop in a rotating background, we need to transform the boundary coordinates of the metric Eq.(\ref{eq201}). There are two ways to introduce rotation. One approach, referred to as ``global rotation", is usually used in the lattice \cite{Braguta:2021ucr, Braguta:2021jgn} and effective field theories. The other, referred to as ``local Lorentz transformation", is employed in some holographic modeling \cite{BravoGaete:2017dso, Braga:2023qej, Zhang:2023psy, Chen:2022obe}. Taking an global rotation   $\phi\rightarrow\phi+\omega t$ on Eq.(\ref{eq201})
\begin{align}
\label{eq21}
    ds^{2} & =\frac{R^{2}}{w^{2}}h\Big[-(f-\omega^{2}l^{2})dt^{2}+l^{2}d\phi^{2}+2\omega l^{2} dtd\phi+dl^{2}+dz^{2}+\frac{1}{f}dw^{2}\Big]\nonumber\\
    &=\frac{R^{2}}{w^{2}}h\bigg\{-[f-\omega^{2}(x^{2}+y^{2})]dt^{2}+2\omega(xdy-ydx)dt+dx^{2} +dy^{2}+dz^{2}+\frac{1}{f}dw^{2}\bigg\},
\end{align} 
here $\omega$ is the angular velocity. The Hawking temperature is
\begin{equation}
\label{eq22}
    T=\frac{1}{\pi w_t}.
\end{equation}
Taking a local Lorentz transformation
\begin{equation}
  t\rightarrow\frac{1}{\sqrt{1-(\omega l_{0})^{2}}}(t+\omega l_{0}^{2}\phi),\quad 
  \phi\rightarrow\frac{1}{\sqrt{1-(\omega l_{0})^{2}}}(\phi+\omega t),  
\end{equation}
here $l_0$ is the spatial distance to the rotation axis. The resulting metric is given by
\begin{align}
\label{eq23}
    ds^{2} &=\frac{R^{2}}{w^{2}}h\bigg\{\frac{1}{1-(\omega l_{0})^{2}}[(-f_l+\omega^{2}l^{2})dt^{2}+(l^{2}-\omega^{2}l_{0}^{4}f_l)d\phi^{2}+2\omega(l^{2}-l_{0}^{2}f_l)dt d\phi] \nonumber \\
    &\quad+dl^{2}+dz^{2}+\frac{1}{f_l}dw^{2}\bigg\}.
\end{align}
For the convenience of later distinction, we use $w_{tl}$ to represent the position of the horizon in the local rotating background. Here $f_l=1-w^4/w_{tl}^4$ with $w_{tl}$ related to the Hawking temperature under a local rotation via
\begin{equation}
\label{eq24}
T=\frac{\sqrt{1-\omega^2l_0^2}}{\pi w_{tl}}.
\end{equation}
We shall explore both transformations for the rest of this note. 

\section{String tension in rotating background}\label{sec3}

The energy of heavy quarkonium can be extracted from the expectation value of the Wilson loop. In the limit $\mathcal{T}\rightarrow\infty$, we have
\begin{equation}
\label{eq301}
  <W(C)>\sim e^{-i\mathcal{T}E(r)},  
\end{equation}
where $C$ is a rectangular loop living on the boundary of $\rm{AdS}$ space with one side representing the time $\mathcal{T}$ and the other side representing the distance $r$ between the quark-antiquark pair.

It follows from the AdS/QCD conjecture, in the large $N_c$ and large 't Hooft coupling limit, the expectation value of the Wilson loop is related to the Nambu-Goto action
\begin{equation}
\label{eq302}
  <W(C)>\sim e^{iS},  
\end{equation}
where
\begin{equation}
\label{eq303}
  S=-\frac{1}{2\pi\alpha^\prime}\int d^2\sigma \sqrt{-\det G_{\mu \nu }\partial_\alpha X^\mu \partial_\beta X^\nu },  
\end{equation}
evaluated at the solution of the Euler-Lagrange equations that yields the minimum area of the string world-sheet bordered by $C$ on the boundary. Combining Eq.(\ref{eq301}) and Eq.(\ref{eq302}), the energy of heavy quarkonium is
\begin{equation}
\label{eq304}
  E(r)=-\frac{S}{\mathcal{T}},  
\end{equation}
in the large seperation distance, $E(r)\sim \kappa r$. The coefficient $\kappa$ is the string tension.

In order to calculate the string tension, we first need to evaluate the energy of heavy quarkonium. In the absence of rotation, the energy is independent of the location and the orientation of the quarkonium. Taking the static solution ansatz with world-sheet coordinates $\sigma^\alpha=(t,w)$,
\begin{equation}
\label{eq_0}
z=z(w),\quad \phi={\rm{const}}.,\quad l={\rm{const}}.,
\end{equation}
it follows from Eq.(\ref{eq21}) and Eq.(\ref{eq303}) that
\begin{equation}
\label{eq031}
 S =-\frac{1}{2\pi\alpha^{\prime}}\int dt dw\mathcal{L}_{0}, 
\end{equation}
with
\begin{equation}
\label{eq0311}
\mathcal{L}_{0}=\frac{R^{2}}{w^{2}}h\sqrt{f\left(z^{\prime2}+\frac{1}{f}\right)},
\end{equation}
where $z^\prime=\frac{dz}{dw}$. Solving the equation of motion of $\mathcal{L}_{0}$, we have
\begin{equation}
\label{eq032}
    z^{\prime}=\sqrt{\frac{g(w_{0})}{f(g(w)-g(w_{0}))}},
\end{equation}
with 
\begin{equation}
  g(w)=\frac{h^{2}(w)}{w^{4}}f(w),\quad g(w_0)=\frac{h^{2}(w_0)}{w_0^{4}}f(w_0).
\end{equation}
The solution corresponds to a U-shaped world-sheet with the turning point at $w=w_{0}$, where $z^\prime|_{w=w_0}\rightarrow\infty$. Substituting Eq.(\ref{eq032}) into Eq.(\ref{eq031}), we find that
\begin{equation}
\label{eq33_S0}
S=-\frac{\sqrt{\lambda}\mathcal{T}}{\pi }\int_{0}^{w_{0}}dw\frac{h}{w^{2}}\sqrt{\frac{g(w)}{g(w)-g(w_{0})}},
\end{equation}
where $\lambda=\frac{R^4}{\alpha^{\prime2}}$ is 't Hooft coupling constant and the distance between the quark-antiquark pair is 
\begin{equation}
\label{eq033}
r=2\int_{0}^{w_{0}}dw\sqrt{\frac{g(w_{0})}{f(g(w)-g(w_{0}))}},\quad 0<w_{0}<w_{min} .  
\end{equation}

\begin{figure}
    \centering
\includegraphics[scale=0.8]{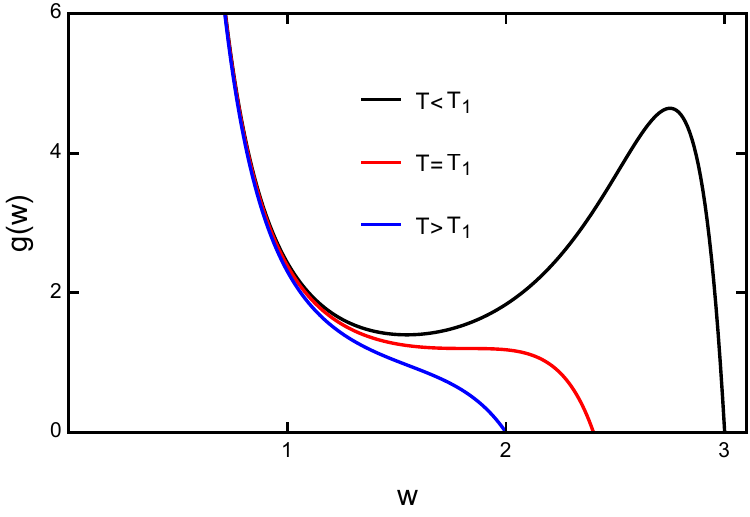}
\caption{$g(w)$ as a function of $w$ at different temperature.}
    \label{fig1}
\end{figure}

Obviously, Eq.(\ref{eq032}) only makes sense when $g(w)>g(w_0)$. As shown in Fig.\ref{fig1}, the graph of $g(w)$ as a function of $w$ is temperature dependent. For temperatures below $T_1\simeq132$ MeV, the function $g(w)$ has a minimum and a maximum at $w_{min}$ and $w_{max}$ respectively. We have 
\begin{equation}
\label{eq0321}
    T_1=\frac{1}{\pi}\sqrt{\frac{c}{\sqrt{27}}},\quad
    w_{min}=w_{t}\sqrt{\frac{2}{\sqrt{3}}\sin\left(\frac{1}{3}\arcsin\frac{T^{2}}{T_{1}^{2}}\right)},\quad
    w_{max}=w_{t}\sqrt{\frac{2}{\sqrt{3}}\sin\left(\frac{\pi}{3}-\frac{1}{3}\arcsin\frac{T^{2}}{T_{1}^{2}}\right)}.
\end{equation}
The RHS of Eq.(\ref{eq032}) is real when $0<w<w_{min}$ and $w_{max}<w<w_t$. There is a wall located at $w=w_{min}$, which means the U-shaped string can not get deeper than $w_{min}$. As stated in \cite{Andreev:2006nw}, this is related to the low-temperature phase. When $w_0$ approaches $w_{min}$ from below, both integrals Eq.(\ref{eq33_S0}) and Eq.(\ref{eq033}) diverge as $\ln\frac{1}{w_{min}-w_0}$ at the upper limit, leading to a linear relationship between $S$ and $r$ for large $r$, i.e. the confining potential. For temperatures above $T_1$, $g(w)$ becomes monotonically decreasing for $w>0$ and Eq.(\ref{eq032}) is valid when $0<w<w_t$, which is related to the deconfined phase. Since we are going to calculate the string tension, we focus on the physics of the low-temperature phase. 

Rotation introduces inhomogeneity and anisotropicity, reflected in the metrics Eq.(\ref{eq21}) and Eq.(\ref{eq23}). The quarkonium energy depends not only on the distance between the quark and the antiquark but also on the location of the center of mass and the direction of the link between the quark and the antiquark. 
Combining Eq.(\ref{eq21}) and Eq.(\ref{eq303}), the Nambu-Goto action for the generic embedding
\begin{equation}
\label{generic}
   \phi=\phi(w),\quad z=z(w),\quad l=l(w), 
\end{equation}
reads
\begin{equation}
\label{eq42}
   S =-\frac{\mathcal{T}}{2\pi\alpha^{\prime}}\int  dw\frac{R^{2}}{w^{2}}h\sqrt{(f-\omega^{2}l^{2})\left(l^{\prime2}+z^{\prime2}+\frac{1}{f}\right)+fl^{2}\phi^{\prime2}}.
\end{equation}
Consequently, the corresponding Euler-Lagrange equations of $\phi$, $z$ and $l$ couples to each other for a nonzero $\omega$. On the other hand, we have proved in \cite{Chen:2023yug} that the solution of the U-shaped world-sheet without rotation Eq.(\ref{eq032}) suffices to evaluate the Nambu-Goto action to the order $\omega^2$\footnote{This statement can be understood simply by considering the minimization of a multivariate function $F(x_1,x_2,...)=F_0(x_1,x_2,...)+\epsilon F_1(x_1,x_2,...)$ with $\epsilon<<1$. To the 0-th order in $\epsilon$, we have solutions, $x_i^{(0)}(i=1,2,...,N)$, as determined by the condition $\frac{\partial F_0}{\partial x_i}=0$. To the order $\epsilon$, the solution is shifted by $\delta x_i=-\epsilon(M^{-1})_{ij}N_j$, where $M$ is the matrix made of the 2nd order partial derivatives $\frac{\partial^2F_0}{\partial x_i\partial x_j}$ and $N$ is the column matrix made of $\frac{\partial F_1}{\partial x_j}$, both evaluated at $x_i^{(0)}$. Substituting $x_i^{(0)}+\delta x_i$ into the target function $F$, we note that to the order $\epsilon$, the shift $\delta x_i$ does not contribute to $F$ because of the 0-th order minimization condition.}. This shortcut will be taken for the rest of the section.

\subsection{Global rotating background}\label{sec3.1}

We consider the heavy quarkonium oriented in the direction of the rotation axis. Substituting the 0-th order solution ansatz Eq.(\ref{eq_0}) to Eq.(\ref{eq42}), we find, to the order of $\omega^2$, that
\begin{equation}
\label{eq31}
 S =-\frac{1}{2\pi\alpha^{\prime}}\int dt dw(\mathcal{L}_{0}+\omega^{2}l_0^{2}\mathcal{L}_{1}), 
\end{equation}
with $l_0$ the distance to the rotation axis and
\begin{equation}
\label{eq311} 
  \mathcal{L}_{1}=-\frac{R^{2}}{2w^{2}}h\sqrt{\frac{1}{f}\left(z^{\prime2}+\frac{1}{f}\right)}.
\end{equation}

Substituting Eq.(\ref{eq032}) to Eq.(\ref{eq31}), the action becomes
\begin{equation}
\label{eq33_S}
S=-\frac{\sqrt{\lambda}\mathcal{T}}{\pi }\int_{0}^{w_{0}}dw\frac{h}{w^{2}}\sqrt{\frac{g(w)}{g(w)-g(w_{0})}}\left(1-\frac{1}{2f}\omega^{2}l_0^{2}\right),
\end{equation}
where $\lambda=\frac{R^4}{\alpha^{\prime 2}}$ is the 't Hooft coupling. Since the above integral diverges at $w=0$ due to the factor $w^{-2}$, we regularize it by subtracting a term $1-\frac{1}{2}\omega^{2}l_0^{2}$ \cite{Andreev:2006ct}. The renormalized action is 
\begin{equation}
\label{eq34}
S_{R}=\frac{\sqrt{\lambda}\mathcal{T}}{\pi}\int_{0}^{w_{0}}dw\frac{1}{w^{2}}\bigg\{h\sqrt{\frac{g(w)}{g(w)-g(w_{0})}}\left(1-\frac{1}{2f}\omega^{2}l_0^{2}\right)-\left(1-\frac{1}{2}\omega^{2}l_0^{2}\right)\bigg\}.
\end{equation}

As $w_0\to w_{\rm {min}}$ from below, both the integrals Eq.(\ref{eq033}) and Eq.(\ref{eq34}) diverge logarithmically, so the long-distance behavior of the potential is contributed by the integration variable $w$ near the upper limit $w_0$ and $w_0$ is close to $w_{\rm{min}}$, where
\begin{equation}
g(w)\simeq g(w_{\rm min})+\frac{1}{2}\frac{d^2g}{dw^2}|_{w=w_{\rm min}}(w-w_{\rm min})^2.
\end{equation}
Introduce $0<\epsilon<<1$ such that $w_0=w_{\rm min}(1-\epsilon)$ and $\epsilon<<\delta<<1$, we have, upon transforming the integration variable to $\xi=(w_0-w)/(w_{\rm min}-w_0)$,
\begin{equation}
\int_{w_{\rm min}(1-\delta)}^{w_{\rm min}(1-\epsilon)}dw\sqrt{\frac{g(w)}{g(w)-g(w_{0})}}\simeq\sqrt{\frac{2g(w_{\rm min})}{\frac{d^2g}{dw^2}|_{w=w_{\rm min}}}}\int_0^\frac{\delta}{\epsilon}\frac{d\xi}{\sqrt{\xi(\xi+2)}}\simeq\sqrt{\frac{2g(w_{\rm min})}{\frac{d^2g}{dw^2}|_{w=w_{\rm min}}}}\ln\frac{\delta}{\epsilon},
\end{equation}
for $\delta>>\epsilon$. Consequently, 
\begin{equation}
   E\simeq\frac{\sqrt{\lambda}}{\pi}\frac{h(w_{min})}{w_{min}^{2}}\sqrt{\frac{g(w_{min})}{\frac{1}{2}\frac{\partial^{2}g}{\partial w^{2}}|_{w=w_{min}}}}\left(1-\frac{1}{2f(w_{min})}\omega^{2}l^{2}\right)\ln\frac{w_{min}}{w_{min}-w_{0}}, 
\end{equation}
in the large distance limit, Eq.(\ref{eq033}) becomes
\begin{equation}
\label{eq35}
  r\simeq 2\frac{h(w_{min})}{w_{min}^{2}\sqrt{\frac{1}{2}\frac{\partial^{2}g}{\partial w^{2}}|_{w=w_{min}}}}\ln\frac{w_{min}}{w_{min}-w_{0}}.  
\end{equation}
A linear potential emerges with the string tension given by
\begin{equation}
 \kappa_g=\frac{\sqrt{\lambda}}{2\pi}\sqrt{g(w_{min})}\left(1-\frac{1}{2f(w_{min})}\omega^{2}l_0^{2}\right),
\end{equation}
combining Eq.(\ref{eq22}) and Eq.(\ref{eq0321}), we have
\begin{equation}
 \kappa_g=\frac{\sqrt{\lambda}\pi T^{2}}{2b}\sqrt{1-b^{2}}e^{\frac{3\sqrt{3}bT_{1}^{2}}{2T^{2}}}\Big[1-\frac{1}{2(1-b^{2})}\omega^{2}l_0^{2}\Big],
\label{longitudinal}
\end{equation}
with
\begin{equation}
b=\frac{w_{\rm min}^2}{w_t^2}=\frac{2}{\sqrt{3}}\sin\left(\frac{1}{3}\arcsin\frac{T^{2}}{T_{1}^{2}}\right).
\label{b}
\end{equation}
If we set $\omega=0$, the result is the same as \cite{Andreev:2006nw}. Obviously, the string tension decreases with increasing angular velocity.
 
The string tension at zero temperature is 
\begin{equation}
\kappa_{g0}=\kappa_0\left(1-\frac{1}{2}\omega^{2}l_0^{2}\right),
\end{equation}
with $\kappa_0=\frac{ce\sqrt{\lambda}}{4\pi}$ the tension without rotation \cite{Andreev:2006ct}.

In the global rotation background, we may also calculate the $\omega^2$ correction of the string tension for a quarkonium at an arbitrary location with an arbitrary orientation of the link between quark and antiquark. As a simple example, we consider a quarkonium to be symmetric with respect to the rotation axis. At $\omega=0$, the U-string solution reads
\begin{equation}
x(w)=\pm \int_w^{w_0}dw^{\prime}\sqrt{\frac{g(w_{0})}{f(w^{\prime})(g(w^{\prime})-g(w_{0}))}},
\end{equation}
with $y=0$ and $z={\rm const.}$. If we expand the Nambu-Goto action to $\omega^2$ term $S=S_0+\omega^2S_1$, then
\begin{align}
S_1 &= \frac{\sqrt{\lambda}\mathcal{T}}{\pi}\int_0^{w_0}dw\frac{hx^2(w)}{2w^2f}\sqrt{\frac{g(w)}{g(w)-g(w_{0})}}\nonumber\\
&= \frac{\sqrt{\lambda }\mathcal{T}}{\pi}\int_0^{w_0}dw\frac{h}{2w^2f}\sqrt{\frac{g(w)}{g(w)-g(w_{0})}}\left[\int_w^{w_0}dw^{\prime}\sqrt{\frac{g(w_{0})}{f(w^{\prime})(g(w^{\prime})-g(w_{0}))}}\right]^2.
\end{align}
Like the previous case with the link between quark and antiquark parallel to the rotation axis, the large r behavior corresponds to $w_0$ close to $w_{\rm min}$ and the integration variable $w$ close to the upper limit $w_0$. With the aid of the integration formula
\begin{equation}
\int_0^\frac{\delta}{\epsilon}\frac{d\xi}{\sqrt{\xi(\xi+2)}}\left[\int_0^\xi\frac{d\xi'}{\sqrt{\xi'(\xi'+2)}}\right]^2=\frac{1}{3}\left[\int_0^\frac{\delta}{\epsilon}\frac{d\xi}{\sqrt{\xi(\xi+2)}}\right]^3\simeq\frac{1}{3}\ln^3\frac{\delta}{\epsilon},
\end{equation}
we end up with
\begin{equation}
\label{eq36}
S_1=\frac{\sqrt{\lambda}\mathcal{T}h(w_{\rm min})g^{\frac{3}{2}}(w_{\rm min})}{6\pi f^2(w_{\rm min})w_{\rm min}^2\left(\frac{1}{2}\frac{\partial^{2}g}{\partial w^{2}}|_{w=w_{min}}\right)^{\frac{3}{2}}}\ln^3\frac{w_{\rm min}}{w_{\rm min}-w_0}.
\end{equation}
The relation between the quark-antiquark distance Eq.(\ref{eq033}) and $w_0$ remains intact. Combining Eq.(\ref{eq304}), Eq.(\ref{eq35}) and Eq.(\ref{eq36}), we get the modified string tension
\begin{equation}
\kappa_g=\frac{\sqrt{\lambda}\pi T^{2}}{2b}\sqrt{1-b^{2}}e^{\frac{3\sqrt{3}bT_{1}^{2}}{2T^{2}}}\left[1-\frac{1}{24(1-b^{2})}\omega^{2}r^{2}\right].
\label{transverse}
\end{equation}
The general expression of the string tension is derived in Appendix A and we find that
\begin{equation}
\kappa_g=\frac{\sqrt{\lambda}\pi T^{2}}{2b}\sqrt{1-b^{2}}e^{\frac{3\sqrt{3}bT_{1}^{2}}{2T^{2}}}\bigg\{1-\frac{\omega^2}{2(1-b^{2})}\left[(\vec m\times\vec X)^2-(\vec n\cdot\vec m\times\vec X)^2+\frac{1}{12}r^2(\vec m\times\vec n)^2\right]\bigg\},
\label{general}
\end{equation}
where $\vec X$ is the center coordinates of the quarkonium, $\vec n$ is the unit vector in the direction of the link between the quark and antiquark, and $\vec m$ is the unit vector of the rotation axis. 

For QCD, $\sqrt{\kappa_g}$ is estimated to be around 465 MeV, for the angular velocity estimated for the off-central collision, $\omega\simeq 7$ MeV \cite{Edwards:1997xf}. It follows from Eq.(\ref{transverse}) that there is a region of $r$, where the linear potential approximates well with a small nonlinear correction by the rotation. In all cases, the string tension is reduced by the rotation.

\subsection{Local rotating background}\label{sec3.1}
Using the same method as the above subsection, we can get the Nambu-Goto action in the local rotating background with Eq.(\ref{eq23}),
\begin{equation}
   S_l=-\frac{1}{2\pi\alpha^{\prime}}\int dt dw(\mathcal{L}^\prime_{0}+\omega^{2}l_{0}^{2}\mathcal{L}^\prime_{1}), 
\end{equation}
with
\begin{equation}
\label{eq370}
  \mathcal{L}^\prime_{0}=\frac{R^{2}}{w^{2}}h\sqrt{f_l\left(z^{\prime2}+\frac{1}{f_l}\right)},\quad 
  \mathcal{L}_{1}^{\prime}=\frac{1}{2}\frac{R^{2}}{w^{2}}h(f_l-1)\sqrt{\frac{1}{f_l}\left(z^{\prime2}+\frac{1}{f_l}\right)}.
\end{equation}
The large distance behavior of the energy is 
\begin{equation}
\label{eq37}
 E_l\sim\frac{\sqrt{\lambda}}{\pi}\frac{h(w^\prime_{min})}{w^{\prime 2}_{min}}\sqrt{\frac{g(w^\prime_{min})}{\frac{1}{2}\frac{\partial^{2}g}{\partial w^{2}}|_{w=w^\prime_{min}}}}\left(1+\frac{f_l(w^\prime_{min})-1}{2f_l(w^\prime_{min})}\omega^{2}l_0^{2}\right)\ln\frac{w^\prime_{min}}{w^\prime_{min}-w_{0}}.    
\end{equation}
The difference between $w_{min}$ and $w^\prime_{min}$ is that $w_t$ in Eq.(\ref{eq0321}) is replaced by $w_{tl}$. Since the form of $\mathcal{L}^\prime_{0}$ in Eq.(\ref{eq370}) is the same as $\mathcal{L}_0$ in Eq.(\ref{eq0311}), the distance between the quark-antiquark pair can be obtained by substituting $w_{min}$ for $w^\prime_{min}$. The string tension in the local rotating background is
\begin{equation}
\label{eq38}
\kappa_l=\frac{\sqrt{\lambda}}{2\pi}\sqrt{g(w^\prime_{min})}\left(1+\frac{f_l(w^\prime_{min})-1}{2f_l(w^\prime_{min})}\omega^{2}l_0^{2}\right).  
\end{equation}
Next, we compare the difference between the string tension in the global rotating
background and in the local rotating background. Combining Eq.{\ref{eq22}} and Eq.(\ref{eq24}) and keeping the temperature the same, we get the relation between $w_{tl}$ and $w_t$ to the order of $\omega^2$
\begin{equation}
  w_{tl}=\left(1-\frac{1}{2}\omega^{2}l_{0}^{2}\right)w_{t}.  
\end{equation}
Substituting it into Eq.(\ref{eq38}), we have
\begin{equation}
 \kappa_l=\frac{\sqrt{\lambda}\pi T^{2}}{2b}\sqrt{1-b^{2}}e^{\frac{3\sqrt{3}bT_{1}^{2}}{2T^{2}}}\bigg\{1-\Big[\frac{b^{2}}{2(1-b^{2})}+\frac{3\sqrt{3}bT_{1}^{2}}{2T^{2}}-1\Big]\omega^{2}l_{0}^{2}\bigg\}.  
\end{equation}
The conclusion is the same as that of the global rotating background. The string tension decreases with the increase in angular velocity. The string tension in the local rotating background at zero temperature is independent of rotation because of Lorentz invariance.

Strictly speaking, the energy $E$ calculated in section \ref{sec3} is the Helmholtz free energy of the gluon system in the presence of a heavy quarkonium. An alternative string tension can be extracted from the thermodynamic internal energy by adding back the entropy effect. As shown in \cite{Andreev:2006ct}, the entropy contribution without rotation stays insignificant in the confinement phase until the temperature approaches $T_1$. We find this remains the case with rotation and the rotation suppresses the string tension extracted from the internal energy until the temperature gets close to $T_1$, where the entropy dominates and the string tension extracted from the internal energy is enhanced by the rotation. The details of our analysis can be found in Appendix B. 

\section{Polyakov loop in global rotating background}\label{sec4}

There has been intense interest in confinement-deconfinement phase transition temperature, which can be roughly determined from the expectation value of the Polyakov loop. It is zero below transition temperature $T_d$ by construction. In this section, we will investigate the influence of rotation on the transition temperature. At large $N_c$ and large $\lambda$, the thermal average of the Polyakov loop is given by
\begin{equation}
\label{eq41}
    L=e^{-S_q},
\end{equation}
here $S_q$ is the Nambu-Goto action of the world-sheet of a single quark with an imaginary time. Instead of introducing an imaginary angular velocity as in lattice calculations \cite{Braguta:2021ucr, Braguta:2021jgn}, we start with the Nambu-Goto action $S$ with real-time and make the Wick rotation
\begin{equation}
\mathcal{T}\to-\frac{i}{T},\qquad S\to iS_q,
\label{Wick}
\end{equation}
with $\mathcal{T}$ the real-time extension of a static world sheet.

\subsection{Global rotating background}\label{sec41}

We take the same world-sheet coordinates $\sigma^\alpha=(t,w)$ as for the heavy quark potential. As the world-sheet of the single quark extends to the horizon, where the shortcut solution taken for the string tension gives rise to a logarithmic divergence in the $\omega^2$ term of the Nambu-Goto action and we have to explore the generic static embedding Eq.(\ref{generic})
We focus on the small $\omega$ approximation for analytical tractability. By analyzing the equations of motion, we find that the following ansatz is reasonable
\begin{equation}
	\begin{split}
    \label{eq43}
	& l=l_{0}+\omega^{2}l_{1}(w),
	\\& \phi=\phi_{0}+\omega\phi_{1}(w),
	\\& z=z_{0}+\omega^{2}z_{1}(w),
	\end{split}
\end{equation}
where $l_0$, $\phi_0$ and $z_0$ are constants. Substituting Eq.(\ref{eq43}) into Eq.(\ref{eq42}), the Lagrangian of the Nambu-Goto action is
\begin{align}
\label{Lagrangian}
\mathcal{L} & =\frac{R^{2}}{w^{2}}h\sqrt{[f-\omega^{2}(l_{0}+\omega^{2}l_{1})^{2}]\left(\omega^{4}l_{1}^{\prime2}+\omega^{4}z_{1}^{\prime2}+\frac{1}{f}\right)+f(l_{0}+\omega^{2}l_{1})^{2}\omega^{2}\phi_{1}^{\prime2}}\nonumber\\&=\mathcal{L}_0+\omega^2 l_0^2 \mathcal{L}_1+\cdots,
\end{align}
with
\begin{equation}
  \label{Lagrangian01}
   \mathcal{L}_{0}=\frac{R^{2}}{w^{2}}h,\quad \mathcal{L}_{1}=-\frac{R^{2}}{2w^{2}}h\left(\frac{1}{f}-f\phi_{1}^{\prime2}\right). 
\end{equation}
Here only leading order terms in $\omega$ are maintained in Eq.(\ref{Lagrangian}). According to Eq.(\ref{Lagrangian01}), we need to determine the form of $\phi_1^\prime$. Since the Lagrangian $\mathcal{L}$ does not contain $\phi_1$, the equation of motion to the order $\omega^2$ becomes
\begin{equation}
\frac{\partial\mathcal{L}_1}{\partial\phi_{1}^{\prime}}=C,
\end{equation}
where $C$ is a constant. After a simple calculation, we have
\begin{equation}
   \phi_{1}^{\prime}=C\frac{w^{2}}{R^{2}hf}, 
\end{equation}
substituting it to Eq.\ref{Lagrangian01}
\begin{equation}
   \label{Lagrangian1}
   \mathcal{L}_{1}=-\frac{R^{2}h}{2w^{2}f}\left(1-C^{2}\frac{w^{4}}{R^{4}h^{2}}\right).
\end{equation}
To eliminate the singularity at $f(w_{t})=0$, the constant can be set to
\begin{equation}
  \label{constant}
   C=\frac{R^{2}}{w_{t}^{2}}h(w_{t}), 
\end{equation}
then, the solution is 
\begin{equation}
  \label{solution}
\phi_{1}^{\prime}=\frac{w^{2}h(w_{t})}{w_{t}^{2}hf}. 
\end{equation}
Combining Eq.(\ref{Lagrangian}), Eq.(\ref{Lagrangian01}) and Eq.(\ref{solution}), and making the Wick rotation Eq.(\ref{Wick}), the action contributing to the Polyakov loop Eq.(\ref{eq41}) reads
\begin{equation}
   S_q=\frac{\sqrt{\lambda}}{2\pi T}\int_{0}^{w_{t}}dw\frac{h}{w^{2}}\left[1-\omega^{2}l_{0}^{2}\frac{1}{2f}\left(1-\frac{w^{4}h^{2}(w_{t})}{w_{t}^{4}h^{2}}\right)\right],
\end{equation}
the integral is divergent at $w=0$. Similarly, we regularize it by subtracting a term $1-\frac{1}{2}\omega^{2}l_0^{2}$ and obtain the renormalized action
\begin{align}
  \label{eq44}
S_{qR}&=\frac{\sqrt{\lambda}}{2\pi T}\int_{0}^{w_{t}}dw\frac{1}{w^{2}}\left[h-\frac{\omega^{2}l_{0}^{2}}{2f}h\left(1-\frac{h^{2}(w_{t})}{h^{2}}\frac{w^{4}}{w_{t}^{4}}\right)-1+\frac{1}{2}\omega^{2}l_{0}^{2}\right]
\nonumber\\&=S_0-\omega^2 l_0^2 S_{1g}+\cdots,
\end{align}
with
\begin{equation}
\label{eq45}
S_{0}=\frac{\sqrt{\lambda}}{2\pi T}\int_{0}^{w_{t}}dw\frac{h-1}{w^{2}},
\end{equation}

\begin{equation}
S_{1g}=\frac{\sqrt{\lambda}}{4\pi T}\int_{0}^{w_{t}}dw\frac{1}{w^{2}}\left[\frac{h}{f}\left(1-\frac{h^{2}(w_{t})}{h^{2}}\frac{w^{4}}{w_{t}^{4}}\right)-1\right].
\end{equation}
Substituting Eq.(\ref{eq44}) to Eq.(\ref{eq41}) and maintaining the $\omega^2$ term, we get the Polyakov loop in the global rotating background
\begin{equation}
\label{eq46}
L_g =e^{-(S_{0}-\omega^{2}l_{0}^{2}S_{1g})}=L_{0}+\omega^{2}l_{0}^{2}L_{1g}+\cdots,
\end{equation}
with
\begin{equation}
\label{eq47}
   L_{0}=e^{-S_{0}},\quad L_{1g}  =S_{1g}e^{-S_{0}},
\end{equation}
where $L_0$ is the Polyakov loop without rotation, which is the same as the result in \cite{Andreev:2009zk}.

\begin{figure}[htbp]
 \centering
 \begin{minipage}{0.49\linewidth}
    \centering
    \includegraphics[width=0.9\linewidth]{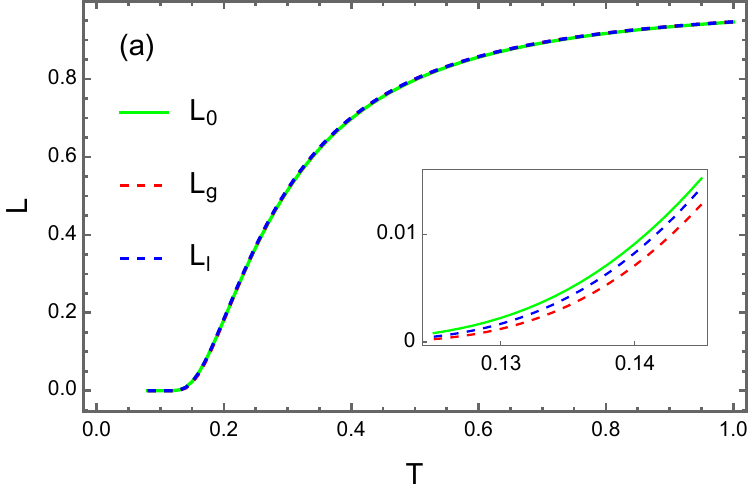}
  \end{minipage}
  \begin{minipage}{0.49\linewidth}
    \centering
    \includegraphics[width=0.9\linewidth]{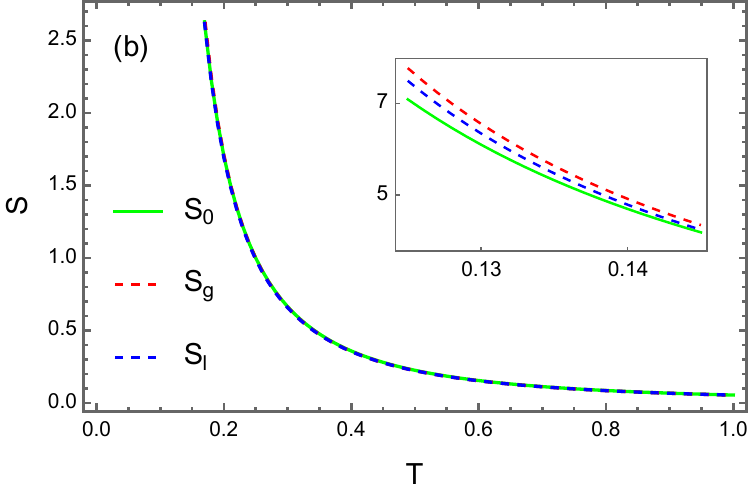}
  \end{minipage}
    \caption{The Polyakov loop $L$ (left) and  the renormalized action $S$ (right) as a function of temperature $T$. Here we set $\sqrt{\lambda}=2.4, \omega l_0=0.2$. The unit of $T$ is $\rm{GeV}$.}
\label{fig2}
\end{figure}

The Polyakov loop in the absence of rotation, $L_0$, and that with rotation up to the order of $\omega^2$, $L_g$, are shown in Figure \ref{fig2} (a) with the renormalized action $S$ displayed in Figure \ref{fig2} (b), where the equations underlying the local rotation frame (blue curves) will be derived later. As a limitation of the holographic formulation \cite{Andreev:2009zk} employed in this work, the deconfinement transition is a crossover in the temperature dependence of the Polyakov loop instead of a discontinuity. Below $T=0.1$ GeV, $L_0$ and $L_g$ are essentially zero ($L_0=4.8\times10^{-10}$ at $T=0.1$ GeV) and rise rapidly to nonzero values in the vicinity of 0.1-0.15 GeV. We define the phase transition temperature $T_d$ as the temperature where the curvature (the reciprocal of the radius of the osculating circle) of the profile $L(T)$ 
\begin{equation}
\label{curvature}
\mathcal{R}(T)=\frac{|\frac{d^2L}{dT^2}|}{\left[1+\left(\frac{dL}{dT}\right)^2\right]^{\frac{3}{2}}},
\end{equation}
is peaked and determine its shift under a small angular velocity. We have
\begin{equation}
\frac{d\mathcal{R}}{dT}\Big|_{T=T_d}=0, \quad \frac{d^2\mathcal{R}}
{dT^2}\Big|_{T=T_d}<0.
\end{equation}
Under a slow rotation
\begin{equation}
\label{curvature_1}
\mathcal{R}(T)\simeq \mathcal{R}_0(T)+\omega^2l_0^2\mathcal{R}_1(T),
\end{equation}
with $\mathcal{R}_0(T)$ the curvature in the absence of rotation which is peaked at $T_d^{(0)}$. The detailed expression of the curvature can be found in Appendix C. It follows that
\begin{equation}
\label{deltaT}
\delta T=T_d-T_d^{(0)}=-\frac{\frac{d\mathcal{R}_1}{dT}}{\frac{d^2\mathcal{R}_0}{dT^2}}\Big\vert_{T=T_d^{(0)}}\omega^2l_0^2=\gamma\omega^2l_0^2,
\end{equation}

\begin{figure}[htbp]
 \centering
 \begin{minipage}{0.49\linewidth}
    \centering
    \includegraphics[width=1\linewidth]{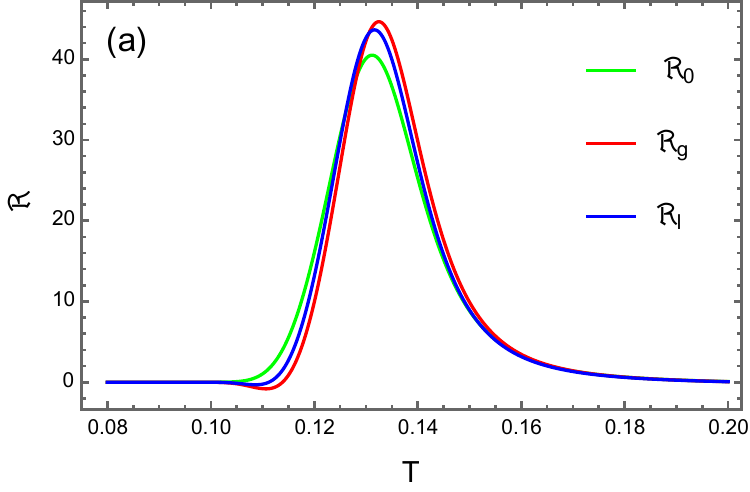}
  \end{minipage}
  \begin{minipage}{0.49\linewidth}
    \centering
    \includegraphics[width=1\linewidth]{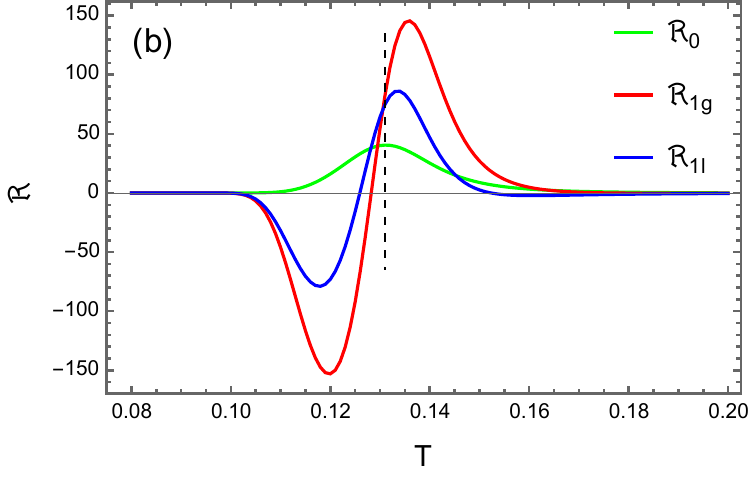}
  \end{minipage}
    \caption{The curvature $\mathcal{R}$ (left) and the curvature without rotation $\mathcal{R}_0$ and the correction term of the curvature $\mathcal{R}_1$ (right) as a function of temperature $T$, where the vertical dashed line in the right panel highlights the location of the curvature peak in the absence of rotation. Here we set $\sqrt{\lambda}=2.4, \omega l_0=0.2$. The unit of $T$ is $\rm{GeV}$.}
 \label{fig3}
\end{figure}

The curvature Eq.(\ref{curvature}) with and without rotation are exhibited in Figure \ref{fig3} (a) at $\sqrt{\lambda}=2.4$, $\omega l_0=0.2$. From Figure \ref{fig3} (a), we find the curvature without rotation $\mathcal{R}_0$ is peaked at $T_d^{(0)}=0.13118$ GeV and the curvature in the global rotating background $\mathcal{R}_g$ is peaked at $T_{dg}=0.13249$ GeV, giving rise to a shift $\delta T_{g}=1.31$ MeV. To estimate the coefficient $\gamma$ in the analytic formula (\ref{deltaT}), both $\mathcal{R}_0$ and $\mathcal{R}_1$ are plotted in Figure \ref{fig3} (b), which shows clearly that $\frac{d\mathcal{R}_{1g}}{dT}>0$. 
Since $\frac{d^2\mathcal{R}_0}{dT^2}<0$ at $T_d^{(0)}$, we find that $\gamma>0$, in line with the shift shown in Figure \ref{fig3}(a). Working out the numerical values of the derivatives in Eq.(\ref{deltaT}), we end up with $\gamma_g=0.0428$ GeV in the global rotating background. If we set $\omega l_0=0.2$, then $\delta T_g=1.71 $ MeV. The difference from the shift read off from Figure \ref{fig3} (a) can be attributed to higher powers in $\omega l_0$ in the presence of large magnitude of $\mathcal{R}_1$ relative to $\mathcal{R}_0$. As our calculation is up to $\omega^2 l_0^2$, only the leading order correction $\delta T_g=0.0428 \omega^2 l_0^2$ GeV is robust. 

The values of the curvature peak without rotation $T_d^{(0)}$ and the shift calculated with formula (\ref{deltaT}) in the global rotating background $\delta T_g$, at different 't Hooft coupling constant can be seen in table \ref{tabel1}, where the value $\sqrt{\lambda}=0.94$ was used in \cite{Andreev:2006nw} and the value $\sqrt{\lambda}=1.44$ was used in \cite{Andreev:2009zk}. From table \ref{tabel1}, we find the shift $\delta T_g$ decreases with increasing 't Hooft coupling constant. 

\begin{table}[htbp]
    \centering
    \begin{tabular}{|c|c|c|c|c|c|}
    \hline
    $\sqrt{\lambda}$ & $T_d^{(0)}$ &  $\delta T_g$ & $\delta T_l$ & $C_g$ & $C_l$ \\ \hline
       0.94 & 104.21 & 5.28 & 3.92 & 0.63 & 0.47
       \\ \hline
       1.44 & 114.18 & 3.29
       & 2.00 & 0.36 & 0.22
       \\ \hline
       2.4  & 131.18 & 1.71 & 0.57 & 0.16 & 0.05     \\ \hline
    \end{tabular}
    \caption{The values of the curvature peak without rotation $T_d^{(0)}$, the shift calculated with formula (\ref{deltaT}) in the global/local rotating background $\delta T_g$, $\delta T_l$, the coefficient in Eq.(\ref{shift}) in the global/local rotating background $C_g$, $C_l$ at different 't Hooft coupling constant. Here we set $\omega l_0=0.2$. The units of $T_d^{(0)}$, $\delta T_g$ and $\delta T_l$ are MeV.}
    \label{tabel1}
\end{table}

To compare our result with the lattice simulation reported in \cite{Braguta:2021jgn}, we have to take the spatial average of the Polyakov loop evaluated before. As only the correction term is coordinate dependent, the spatial average in a cylindrical volume of radius $\rho_0$ around the rotation axis reads
\begin{equation}
<L>=L_0+L_1\omega^2<l_0^2>=L_0+\frac{1}{2}\omega^2\rho_0^2.
\end{equation}
Consequently
\begin{equation}
\label{shift}
\frac{\delta T}{T_d^{(0)}}=\frac{\gamma}{2T_d^{(0)}}v^2\equiv C_i v^2,
\end{equation}
with $v=\omega\rho_0$ the linear velocity on the boundary. The coefficient in Eq.(\ref{shift}) in the global rotating background $C_g$ at different 't Hooft coupling constant are shown in table \ref{tabel1}. We have $C_g=0.63$ when $\sqrt{\lambda}=0.94$, very close to the lattice value of 0.7.

\subsection{Local rotating background}\label{sec42}

In the same way, we get the renormalized action in the local rotating background using the metric Eq.(\ref{eq23})
\begin{equation}
\label{eq421}
S_{qRl}=\frac{\sqrt{\lambda}}{2\pi T}\int_{0}^{w_{tl}}dw\frac{1}{w^{2}}\left[h-\omega^{2}l_{0}^{2}\frac{hw^{4}}{2fw_{tl}^{4}}\left(1-\frac{h^{2}(w_{tl})}{h^{2}}\right)-1\right],
\end{equation}
we need to note that $w_{tl}$ is different to $w_t$ in Eq.(\ref{eq44}). To compare the difference between the Polyakov loop in the global rotating background and in the local rotating background, we change the variable $w$ to $w^\prime=\frac{w}{\sqrt{1-\omega^{2}l_{0}^{2}}}$. Then the upper limit of the integral in Eq.(\ref{eq421}) becomes $w_t=\frac{1}{\pi T}$. We still use $w$ to represent the integral variable below. We have
\begin{equation}
S_{qRl}=S_{0}-\omega^{2}l_{0}^{2}S_{1l}+\cdots,
\end{equation}
with
\begin{equation}
S_{1l}=\frac{\sqrt{\lambda}}{2\pi T}\int_{0}^{w_{t}}dw\frac{hw^{2}}{2fw_{t}^{4}}\left(1-\frac{h^{2}(w_{t})}{h^{2}}\right)+\frac{\sqrt{\lambda}}{4}(e^{\frac{1}{2}cw_{t}^{2}}-1),
\end{equation}
where $S_0$ is the same Eq.(\ref{eq45}). Similarly, the Polyakov loop in the local rotating background is
\begin{equation}
\label{eq422}
L_l=e^{-(S_{0}-\omega^{2}l_{0}^{2}S_{2})}=L_{0}+\omega^{2}l_{0}^{2}L_{1l}+\cdots, 
\end{equation}
with
\begin{equation}
\label{eq423}
   L_{1l}=S_{1l}e^{-S_{0}},
\end{equation}
here $L_0$ is the Polyakov loop without rotation. The Polyakov loop and the curvature $\mathcal{R}_l$ up to $\omega_0^2l_0^2$ as well as the coefficient $\mathcal{R}_{1l}$ for the local rotating frame are depicted in Figures \ref{fig2} and \ref{fig3} in blue lines. As shown in table \ref{tabel1}, the leading order shift of the phase transition temperature of Eq.(\ref{deltaT}) and the coefficient of Eq.(\ref{shift}) in the local rotating background $\delta T_l$, $C_l$, is smaller than that of the global rotation and decrease with increasing 't Hooft coupling constant as well.

\section{Concluding remarks}\label{sec5}

In this paper, we study the effect of rotation on the gluodynamics at strong coupling in terms of gauge/gravity duality. The rotation is introduced by performing a transformation on the static black hole solution. Here two different transformations are considered, which are referred to as ``global rotation'' and ``local Lorentz transformation''. Since the simplistic solution ansatz without rotation ceases to work in the rotating background, we make a small angular velocity approximation and maintain the leading order term in $\omega$. We focus our attention on the string tension underlying quark confinement and the confinement-deconfinement phase transition temperature. 

The string tension is extracted from the heavy quark potential (the free energy in the presence of a heavy quarkonium), which corresponds to a U-shaped world-sheet minimizing the Nambu-Goto action. It is shown that the string tension decreases with increasing angular velocity in all cases examined. This result is consistent with the general expectation that rotation might weaken the confining force.  The confinement-deconfinement transition is characterized by the Polyakov loop, which corresponds to a straight string twisted by rotation towards the AdS bulk. The result shows that the transition temperature increases with increasing angular velocity in the global/local rotation frame, consistent with the lattice simulation result \cite{Braguta:2021jgn}.  However, it is different from other holographic studies \cite{Zhao:2022uxc, Chen:2020ath} that suggest a decrease in the critical temperature of deconfinement with increasing angular velocity. We attribute this discrepancy to differences in the solution ansatz of the Euler-Lagrangian equation of the Nambu-Goto action, which in turn affects the shape of the string.  If we take the world-sheet coordinates of the quark as $\sigma^{\alpha}=(t,w)$ and $\phi$, $z$ and $l$ are constants, which corresponds to a straight string extending from the boundary to the horizon, we will get the same conclusion with others. However, the shape of the string in our paper is not straight, which is defined by Eq.(\ref{solution}). The different shapes of the strings lead to different results.

It is important to note that a decreasing string tension and an increasing critical temperature with rotation may not necessarily contradict each other. The deconfinement transition, being a first-order phase transition, occurs before the string tension vanishes.

Mathematically, our formulation applies to the strong ’t Hooft coupling regime, i.e., $\lambda>>1$. We neglected subleading contributions from fluctuations around the optimal string world-sheet, which are suppressed relative to the leading order by a factor of $(1/\sqrt{\lambda})$. The values of $\lambda$ used in this work may not fully justify the large $\lambda$ approximation. Additionally, Table \ref{tabel1}  suggests that the dependence of the transition temperature on the angular velocity weakens with increasing  $\lambda$. The robustness of our results should be further investigated with more rigorous scrutiny.

In summary, our study provides new insights into the interplay between rotation and gluodynamics at strong coupling, highlighting the importance of the string shape in determining the confinement-deconfinement transition temperature. Future research should explore these effects in greater detail, particularly in the context of varying $\lambda$  and the implications for the string shape ansatz.

\section*{Acknowledgements}
 We thank Yan-Qing Zhao for useful discussions. This work is supported in part by the National Key Research and Development Program of China under Contract No. 2022YFA1604900. This work is also partly supported by the National Natural Science Foundation of China (NSFC) under Grants No. 12435009, and No. 12275104.

\appendix
\section{String tension with arbitrary location and orientation}
\label{appendix A}
In this appendix, we calculate the string tension of the quarkonium with an arbitrary location
and orientation inside a rotating QGP to the order of $\omega^{2}$.

Assuming that the center of the pair is located at
\begin{equation}
\vec {X}=({{l_0}\cos\alpha,{l_0}\sin\alpha,Z}),
\end{equation}
the unit vector along the link between the quark
and antiquark is given by
\begin{equation}
\vec {n}=(\sin\theta \cos\phi,\sin\theta \sin\phi ,\cos\theta).
\end{equation}
We can also define the unit vector $\vec {m}=(0,0,1)$ 
in the direction of the rotation axis.
The coordinate of a point along the link is
\begin{equation}
\vec {x}=\vec {X}+\zeta \vec {n}.
\end{equation}
If we take the world-sheet coordinates as $\sigma^\alpha=(t,w)$, then we can obtain
\begin{equation}
d\vec {x}=\vec {n}\zeta^{\prime}dw.
\end{equation}
It follows from the second line of Eq.(\ref{eq21}) that the induced  metric reads
\begin{align}
	ds^2 &= \frac{R^2}{w^2}h\bigg\{-\left[f-\omega^2\left({l_0}^2+2{l_0}\zeta\sin\theta\cos\beta+\zeta^2\sin^2\theta\right)\right]dt^2\nonumber\\ &- 2\omega b\sin\theta\sin\beta\zeta^\prime dtdr+\left(\zeta^{\prime 2}+\frac{1}{f}\right)dw^2\bigg\},
	\label{worldsheet}
\end{align}
with $\beta=\alpha-\phi$.
Substituting the zeroth order solution
\begin{equation}
\zeta(w)=\pm\int_w^{w_0}dw'\sqrt{\frac{g(w_{0})}{f(g(w')-g(w_{0}))}},
\end{equation}
with $w_0$ determined by
\begin{equation}
r=2\int_0^{w_0}dw^{\prime}\sqrt{\frac{g(w_{0})}{f(g(w^{\prime})-g(w_{0}))}}\simeq   2\frac{h(w_{min})}{w_{min}^{2}\sqrt{\frac{1}{2}\frac{\partial^{2}g}{\partial w^{2}}|_{w=w_{min}}}}\ln\frac{w_{min}}{w_{min}-w_{0}},
\end{equation}
for large $r$.
The determinant of the metric Eq.(\ref{worldsheet}) takes the form
\begin{equation}
	\gamma=-\frac{R^4h^2}{w^4}\left[1-\frac{\omega^2}{f}\left(l_0^2+\zeta^2\sin^2\theta+2l_0\zeta\sin\theta\cos\beta+\frac{g_0}{g}l_0^2\sin^2\theta\sin^2\beta\right)\right]\frac{g(w)}{g(w)-g(w_0)}.
\end{equation}
It follows that the Nambu-Goto action of the U-shaped string to the order $O(\omega^2)$
\begin{align}
	S_U &= -\frac{\mathcal{T}\sqrt{\lambda}}{2\pi}\int_0^{w_0} dw\frac{h}{w^2} \bigg\{\left[1-\frac{\omega^2}{2f}\left(l_0^2+\zeta^2\sin^2\theta+2l_0|\zeta|\sin\theta\cos\beta+\frac{g_0}{g}l_0^2\sin^2\theta\sin^2\beta\right)\right]\nonumber\\
	&\quad+\left[1-\frac{\omega^2}{2f}\left(l_0^2+\zeta^2\sin^2\theta-2l_0|\zeta|\sin\theta\cos\beta+\frac{g_0}{g}l_0^2\sin^2\theta\sin^2\beta\right)\right]\bigg\}\sqrt{\frac{g(w)}{g(w)-g(w_0)}}\nonumber
	\\&= -\frac{\mathcal{T}\sqrt{\lambda}}{\pi}\int_0^{w_0}dw\frac{h}{w^2}\left[1-\frac{\omega^2}{2f}\left(l_0^2+\zeta^2\sin^2\theta-\frac{g_0}{g}l_0^2\sin^2\theta\sin^2\beta\right)\right]\sqrt{\frac{g(w)}{g(w)-g(w_0)}}, 
	\label{U_general} 
\end{align}
where we have separated the branch of $\zeta>0$ from $\zeta<0$ but the difference is cancelled in the end. For large $r$, the turning point $w_0$ is close to $w_{\rm min}$ and the integral is dominated near the upper limit $w_0$. Combining the logics behind (\ref{longitudinal}) and (\ref{transverse}), we obtain the string tension
\begin{equation}
\kappa_g=\frac{\sqrt{\lambda}\pi T^{2}}{2b}\sqrt{1-b^{2}}e^{\frac{3\sqrt{3}bT_{1}^{2}}{2T^{2}}}\bigg\{1-\frac{\omega^2}{2(1-b^{2})}\left[l_0^2(1-\sin^2\theta\sin^2\beta)+\frac{1}{12}r^2\sin^2\theta\right]\bigg\}.
\end{equation}
Using the relationships
\begin{equation}
l_0^2=(\vec e_z\times\vec X)^2 ,\qquad l_0^2\sin^2\theta\sin^2\beta=(\vec n\cdot\vec e_z\times\vec X)^2 ,\qquad \sin^2\theta=(\vec e_z\times \vec n)^2,
\label{eq{general}}
\end{equation}
where $\vec e_z=\vec m$, we can easily derive the relationship (\ref{general}).

\section{String tension extracted from internal energy}
\label{appendix B}

In terms of the Helmholtz free energy $E$ computed in section \ref{sec3}, the entropy of the gluon system in the presence of a heavy quarkonium reads
\begin{equation}
S=-\frac{\partial E}{\partial T},
\end{equation}
and the thermodynamic relation $E=U-TS$ gives rise to the internal energy $U$.
As both $E$ and $U$ shows linear potential for large separation between the quark and antiquark, the string tension defined by the internal energy reads
\begin{equation}
\kappa^U=\kappa-T\frac{\partial\kappa}{\partial T},
\end{equation}
with $\kappa$ the string tension calculated in section \ref{sec3}. In terms of the leading order of the rotation contribution $\delta\kappa$, the $O(\omega^2l_0^2)$ term in $\kappa$, the leading order rotation contribution in $\kappa^U$ is given by
\begin{equation}
\delta\kappa^U=\delta\kappa-T\frac{\partial\delta\kappa}{\partial T}.
\label{Uansatz}
\end{equation}

For the string tension $\kappa$ corresponding to the heavy quark and antiquark separated in the direction of the rotation axis in the global rotation frame, it follows from Eq.(\ref{longitudinal}) that
\begin{equation}
\delta\kappa=-\frac{3\pi e\sqrt{3\lambda}T_1^2}{8}\omega^2\l_0^2\left[1+O\left(\frac{T^4}{T_1^4}\right)\right],
\end{equation}
for $T<<T_1$, and the entropy term in (\ref{Uansatz})is suppressed relative to $\kappa$ by $O(T^4/T_1^4)$. Consequently, $\delta\kappa^U\simeq\delta\kappa<0$ and the rotation reduces $\kappa^U$ as well. 

As $T\to T_1$,
\begin{equation}
\delta\kappa\to-\frac{3\pi}{4}\sqrt{\frac{\lambda}{2}}e^\frac{3}{2}T_1^2\omega^2l_0^2.
\end{equation}
The entropy contribution dominates because of the divergence of the derivative
\begin{equation}
\frac{db}{dT}\simeq\frac{1}{3\sqrt{T_1(T_1-T)}}.
\end{equation}
Consequently, 
\begin{equation}
\delta\kappa^U\simeq -\sqrt{\frac{T_1}{3(T_1-T)}}\delta\kappa>0,
\end{equation}
and the rotation enhances $\kappa^U$. 

It is straightforward to generalize this analysis to all cases of $\kappa$ considered in section \ref{sec3}.

\section{Curvature formula}
\label{appendix C}

 The curvature without rotation, i.e. the first term on RHS of Eq.(\ref{curvature_1}), is
\begin{equation}
\mathcal{R}_{0}=\frac{|\frac{d^2L_0}{dT^2}|}{\left[1+\left(\frac{dL_0}{dT}\right)^2\right]^{\frac{3}{2}}}=\frac{|e^{-S_{0}}\left[\left(\frac{dS_{0}}{dT}\right)^{2}-\frac{d^{2}S_{0}}{dT^{2}}\right]|}{\left[1+e^{-2S_{0}}\left(\frac{dS_{0}}{dT}\right)^{2}\right]^{\frac{3}{2}}},
\end{equation}
where
\begin{equation}
\frac{dS_{0}}{dT}=-\frac{1}{2T^{2}}\sqrt{\frac{\lambda c}{2\pi}}Erfi(\sqrt{s}),
\end{equation}

\begin{equation}
\frac{d^{2}S_{0}}{dT^{2}}	=\frac{1}{T^{3}}\sqrt{\frac{\lambda c}{2\pi}}Erfi(\sqrt{s})+c\sqrt{\lambda}\frac{e^{s}}{2\pi^{2}T^{4}},
\end{equation}
with $s=\frac{c}{2\pi^{2}T^{2}}$. $Erfi$ is a virtual error function.

Combining Eq.(\ref{eq46}) and Eq.(\ref{curvature}), the curvature in the global rotating background is
\begin{equation}
\mathcal{R}_g=\frac{|L_g^{\prime\prime}|}{(1+L_g^{\prime2})^{\frac{3}{2}}},
\end{equation}
with
\begin{equation}
   \frac{dL_g}{dT}=-e^{-S_{0}}\frac{dS_{0}}{dT}+\omega^{2}l_{0}^{2}e^{-S_{0}}\left(\frac{dS_{1g}}{dT}-S_{1g}\frac{dS_{0}}{dT}\right),
\end{equation}

\begin{equation}
  \frac{d^{2}L_g}{dT^{2}}=e^{-S_{0}}\left[\left(\frac{dS_{0}}{dT}\right)^{2}-\frac{d^{2}S_{0}}{dT^{2}}\right](1+\omega^{2}l_{0}^{2}S_{1g})+\omega^{2}l_{0}^{2}e^{-S_{0}}\left(\frac{d^{2}S_{1g}}{dT^{2}}-2\frac{dS_{0}}{dT}\frac{dS_{1g}}{dT}\right),  
\end{equation}

\begin{equation}
 \frac{dS_{1g}}{dT}=-\frac{\sqrt{\lambda}}{4}\frac{ce^{s}}{\pi^{2}T^{3}}\int_{0}^{1}dx\bigg\{\frac{1+2x^{2}-x^{4}}{1-x^{4}}\sinh[s(x^{2}-1)]+\frac{1-x^{2}}{1+x^{2}}\cosh[s(x^{2}-1)]\bigg\},  
\end{equation}

\begin{align}
\frac{d^{2}S_{1g}}{dT^{2}}&=\frac{\sqrt{\lambda}}{4}\frac{c^{2}e^{s}}{\pi^{4}T^{6}}\int_{0}^{1}dxx^{2}\bigg\{\frac{5-4x^{2}+x^{4}}{1-x^{4}}\sinh[s(x^{2}-1)]+\frac{x^{2}-3}{1+x^{2}}\cosh[s(x^{2}-1)]\bigg\}
\nonumber \\
&\quad  +\frac{\sqrt{\lambda}}{4}\frac{3ce^{s}}{\pi^{2}T^{4}}\int_{0}^{1}dx\bigg\{\frac{1+2x^{2}-x^{4}}{1-x^{4}}\sinh[s(x^{2}-1)]+\frac{1-x^{2}}{1+x^{2}}\cosh[s(x^{2}-1)]\bigg\},
\end{align}
keeping the curvature to $\omega^2$ term, we find
\begin{align}
\mathcal{R}_g&=\mathcal{R}_0+\omega^{2}l_{0}^{2}\frac{e^{-S_{0}}}{\left[1+e^{-2S_{0}}\left(\frac{dS_{0}}{dT}\right)^{2}\right]^{\frac{3}{2}}}\bigg\{S_{1g}\left[\left(\frac{dS_{0}}{dT}\right)^{2}-\frac{d^{2}S_{0}}{dT^{2}}\right]+\left(\frac{d^{2}S_{1g}}{dT^{2}}-2\frac{dS_{0}}{dT}\frac{dS_{1g}}{dT}\right) \nonumber \\
&\quad+\frac{3e^{-2S_{0}}}{1+e^{-2S_{0}}\left(\frac{dS_{0}}{dT}\right)^{2}}\frac{dS_{0}}{dT}\left[\left(\frac{dS_{0}}{dT}\right)^{2}-\frac{d^{2}S_{0}}{dT^{2}}\right]\left(\frac{dS_{1g}}{dT}-S_{1g}\frac{dS_{0}}{dT}\right)\bigg\},
\end{align}
where the coefficient of $\omega^2 l_0^2$ is $\mathcal{R}_1(T)$ in (\ref{curvature_1}) for the global rotation.

Similarly, combining Eq.(\ref{curvature}) and Eq.(\ref{eq422}), the curvature in the local rotating background at low temperature is
\begin{align}
\mathcal{R}_l&=\mathcal{R}_0+\omega^{2}l_{0}^{2}\frac{e^{-S_{0}}}{\left[1+e^{-2S_{0}}\left(\frac{dS_{0}}{dT}\right)^{2}\right]^{\frac{3}{2}}}\bigg\{S_{1l}\left[\left(\frac{dS_{0}}{dT}\right)^{2}-\frac{d^{2}S_{0}}{dT^{2}}\right]+\left(\frac{d^{2}S_{1l}}{dT^{2}}-2\frac{dS_{0}}{dT}\frac{dS_{1l}}{dT}\right) \nonumber \\
&\quad+\frac{3e^{-2S_{0}}}{1+e^{-2S_{0}}\left(\frac{dS_{0}}{dT}\right)^{2}}\frac{dS_{0}}{dT}\left[\left(\frac{dS_{0}}{dT}\right)^{2}-\frac{d^{2}S_{0}}{dT^{2}}\right]\left(\frac{dS_{1l}}{dT}-S_{1l}\frac{dS_{0}}{dT}\right)\bigg\},
\end{align}
with
\begin{equation}
 \frac{dS_{1l}}{dT}=-\frac{\sqrt{\lambda}}{4}\frac{ce^{s}}{\pi^{2}T^{3}}\bigg\{1+\int_{0}^{1}dx\frac{2x^{2}}{x^{2}+1}\left[\frac{1}{1-x^{2}}\sinh[s(x^{2}-1)]-\cosh[s(x^{2}-1)]\right]\bigg\},   
\end{equation}

\begin{align}
 \frac{d^{2}S_{1l}}{dT^{2}}&=\frac{\sqrt{\lambda}}{4}\frac{c^{2}e^{s}}{\pi^{4}T^{6}}\{1-\int_{0}^{1}dx\frac{2x^{2}}{x^{2}+1}\left[\frac{x^{4}-2x^{2}+2}{x^{2}-1}\sinh[s(x^{2}-1)]-2\cosh[s(x^{2}-1)]\right]\bigg\}  \nonumber
 \\&\quad +\frac{\sqrt{\lambda}}{4}\frac{3ce^{s}}{\pi^{2}T^{4}}\bigg\{1+\int_{0}^{1}dx\frac{2x^{2}}{x^{2}+1}\left[\frac{1}{1-x^{2}}\sinh[s(x^{2}-1)]-\cosh[s(x^{2}-1)]\right]\bigg\}.
\end{align}

	
	
	
\bibliographystyle{utphys}
\bibliography{ref}

\end{document}